\documentclass[a4paper,11pt]{article}
\usepackage{latexsym}
\usepackage{amssymb}
\usepackage{amsfonts}
\usepackage{amsmath}
\usepackage{indentfirst}
\usepackage[dvips]{graphicx}
\usepackage{epsfig}
\newtheorem{prop}{Proposition}[section]

\newcommand{\cvd}{\hfill $\blacksquare$\bigskip}

\newtheorem{example}{Example}[section]

\date{}

\author{S. Bilotta\thanks{Dipartimento di Sistemi e Informatica, viale
Morgagni 65, 50134 Firenze, Italy {\tt bilotta@dsi.unifi.it\quad
pinzani@dsi.unifi.it}}\and F. Disanto\thanks{Dipartimento di Scienze Matematiche ed
Informatiche, Pian dei Mantellini, 44, 53100, Siena, Italy \; {\tt
disafili@yahoo.it\quad rinaldi@unisi.it}}\and R. Pinzani$^*$\and S. Rinaldi$^\dag$}

\title{Catalan structures and Catalan pairs}

\begin{document}

\maketitle

\begin{abstract}
A \emph{Catalan pair} is a pair of binary relations $(S,R)$
satisfying certain axioms. These objects are enumerated by the
well-known Catalan numbers, and have been introduced in
\cite{DFPR} with the aim of giving a common language to most of
the structures counted by Catalan numbers. Here, we give a simple
method to pass from the recursive definition of a generic Catalan
structure to the recursive definition of the Catalan pair on the
same structure, thus giving an automatic way to interpret Catalan
structures in terms of Catalan pairs. We apply our method to many
well-known Catalan structures, focusing on the meaning of the
relations $S$ and $R$ in each considered case.
\end{abstract}

\section{Catalan pairs}

Catalan numbers are a very popular sequence of integer numbers,
arising in many combinatorial problems coming out from different
scientific areas, including computer science, computational
biology, and mathematical physics \cite{St1,St2}. Throughout all
the paper, we will refer to any combinatorial structure enumerated
by Catalan numbers as to a {\em Catalan structure}.

{\em Catalan pairs} have been introduced in \cite{DFPR} with the
aim of giving a common language to (almost) all the Catalan
structures. To reach this goal the authors of \cite{DFPR} use an
elementary mathematical tool, the Catalan pair, which is
substantially a pair of binary relations satisfying certain
axioms. They first prove that Catalan pairs are enumerated by
Catalan numbers, according to their size. Their main goal is to
prove that almost all Catalan structures can be interpreted in
terms of Catalan pairs, thus providing a powerful tool to
determine, in automatic way, many bijections between Catalan
structures. Still in \cite{DFPR}, the authors prove several
combinatorial properties of such Catalan pairs, also showing that
they are related to some classes of pattern avoiding posets.

In this paper we carry on the original purpose of \cite{DFPR}, and
attempt at developing a general method to determine a
representation of a given Catalan structure in terms of Catalan
pairs. Our method relies on the observation that most of the
Catalan structures admit a recursive decomposition, which can be
naturally translated onto the two binary relations defining
Catalan pairs. Once we have presented our methodology, in Section
2 we apply it to furnish the interpretation of some of the most
classical Catalan structures in terms of Catalan pairs. In the
final section, we extend our method in order to include some other
Catalan structures.

\subsection{Basic definitions}
In what follows we recall from \cite{DFPR} some basic definitions
of Catalan pairs. Given any set $X$, we denote
$\mathcal{D}=\mathcal{D}(X)$ the \emph{diagonal} of $X$, that is
the relation $\mathcal{D}=\{ (x,x)\; |\; x\in X\}$. Moreover, if
$\theta$ is any binary relation on $X$, we denote by
$\overline{\theta}$ the \emph{symmetrization} of $\theta$, i.e.
the relation $\overline{\theta}=\theta \cup \theta^{-1}$. Given a
set $X$ of cardinality $n$, let $\mathcal{O}(X)$ be the set of
strict order relations on $X$. By definition, this means that
$\theta \in \mathcal{O}(X)$ when $\theta$ is an irreflexive and
transitive binary relation on $X$.

Now let $(S,R)$ be an ordered pair of binary relations on $X$. We
say that $(S,R)$ is a \emph{Catalan pair} on $X$ when the
following axioms are satisfied:
\begin{itemize}
\item[(i)] $S, R\in \mathcal{O}(X)$;

\item[(ii)] $\overline{R} \cup \overline{S}=X^2 \setminus
\mathcal{D}$;

\item[(iii)] $\overline{R} \cap \overline{S}=\emptyset$;

\item[(iv)] $S\circ R\subseteq R$.
\end{itemize}

One can observe that, since $S$ and $R$ are both strict order
relations, the two axioms (ii) and (iii)  can
be explicitly described by saying that, given $x,y\in X$, with $x
\neq y$, exactly one of the following holds: $xSy$, $xRy$, $ySx$,
$yRx$. The axiom (iv) says that if $xSy$ and $yRz$, then $xRz$.

\medskip

Two Catalan pairs $(S_1 ,R_1 )$ and $(S_2 ,R_2 )$ on the (not
necessarily distinct) sets $X_1$ and $X_2$, respectively, are said
to be \emph{isomorphic} when there exists a bijection $\xi$ from
$X_1$ to $X_2$ such that $xS_1 y$ if and only if $\xi (x)S_2 \xi
(y)$ and $xR_1 y$ if and only if $\xi (x)R_2 \xi (y)$. We will
consider Catalan pairs up to an isomorphism and, as a consequence
of this definition, we say that a Catalan pair has \emph{size} $n$
when it is defined on a set $X$ of cardinality $n$. The set of
isomorphism classes of Catalan pairs of size $n$ will be denoted
$\mathcal{C}(n)$. In \cite{DFPR} it is proved, both in an
analytical and in a bijective way, that the number of Catalan
pairs of size $n$ is indeed the $n$th Catalan number.

\section{Combinatorial interpretations of Catalan pairs}\label{instances}

In this section we provide a general method to represent a given
Catalan structure in terms of a Catalan
pair. More precisely, we start from a given Catalan structure $\mathcal{C}$,
for each object of $\mathcal{C}$ we determine a {\em base set} $X_C$, and we recursively define
a pair of binary relations $(S,R)$ on $X_C$. We can prove that $(S,R)$ is indeed a Catalan pair.

\begin{figure}[!htb]
\begin{center}
\epsfig{file=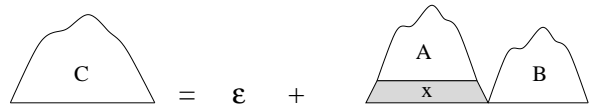,width=3in,clip=} \caption{\small{The
recursive decomposition of Catalan structures illustrated for the
class of Dyck paths.} \label{decc}} \vspace{-15pt}
\end{center}
\end{figure}

We rely on the fact that, using a pretty classical notation, most of the
Catalan structures $\mathcal{C}$ admit a recursive decomposition
as
\begin{equation}\label{deccs}
\mathcal{C}\, = \, \varepsilon \, + \, x \, \mathcal{C} \times  \mathcal{C}
\end{equation}
meaning that, each element $C \in \mathcal{C}$ is the empty object of size zero, or it can be uniquely decomposed
as $C=xAB$, where $x$ is an element of unitary size belonging to the {\em base set} $X_C$, and $A, B \in \mathcal{C}$.
Figure \ref{decc} shows an example of decomposition (\ref{decc}) for the class of {\em Dyck paths}.

From the decomposition (\ref{deccs}), we can recursively define a
base set $X_C$ and a Catalan pair $(S,R)$ on $X_C$ for the Catalan
structure $\mathcal{C}$ in the following way:
\begin{itemize}
\item[$\bullet$] If $C=\varepsilon$ we have $S=\emptyset$ and $R=\emptyset$.\\
\item[$\bullet$] Otherwise, if $C=x A B$, let $(S_A,R_A)$ and
$(S_B,R_B)$ be the Catalan pairs on the objects $A$ and $B$, with
base sets $X_A$, and $X_B$, respectively. Then $X_C$ is composed
by the elements in $X_A$, $X_B$ plus the new element $x$, and
$(S,R)$ is defined as

\begin{eqnarray}\label{definizione}
S&=&S_A \cup S_B \cup \{(a,x): a
\in X_A\}, \\
R&=&R_A \cup R_B \cup \{(a,b): a
\in X_A, b \in X_B\} \cup \{(x,b): b \in
X_B\}\label{definizione2}
\end{eqnarray}
\end{itemize}
The size of $(S,R)$ is then given by the number of elements in $X_C$. We can trivially prove the following statement.

\begin{prop}\label{princ}
The relations $S$ and $R$ defined in (\ref{definizione}) and (\ref{definizione2}) form a Catalan pair on the base set
$X_C$. \cvd
\end{prop}

In the sequel, using Proposition \ref{princ}, we can easily determine the
relations $S$ and $R$ for some known Catalan structures. In
particular, we will take into consideration here some examples from
\cite{St1} and \cite{St2}, involving rather
different combinatorial objects.

According to our method, the relations $S$ and $R$ are recursively defined, but we will be able to present them in an explicit and meaningful way for each
considered case; in particular we will show that each relation describes a certain combinatorial/geometrical relationship among the elements of the base set.

\subsection{Perfect noncrossing matchings and Dyck paths}\label{matchings}

Our first example will be frequently recalled throughout all the
paper. Given a linearly ordered set $A$ of even cardinality, a \emph{perfect
noncrossing matching} of $A$ is a noncrossing partition of $A$
having all the blocks of cardinality 2. A block can be represented by means of an arch joining each couple of points.
There is an obvious bijection between perfect noncrossing matchings and well formed
strings of parentheses. It is known that, the class of perfect
noncrossing matchings is counted by the Catalan numbers according to the number of arches.

Using this representation, we can define
the following relations on the set $X$ of arches of a given
perfect noncrossing matching:

\begin{itemize}
\item for any $x,y\in X$, we say that $xSy$ when $x$ is included
in $y$;

\item for any $x,y\in X$, we say that $xRy$ when $x$ is on the
left of $y$.
\end{itemize}

The reader is invited to check that the above definition yields a
Catalan pair $(S,R)$ on the set $X$.

\bigskip

\begin{figure}[!h]
\begin{center}
\includegraphics[scale=0.44]{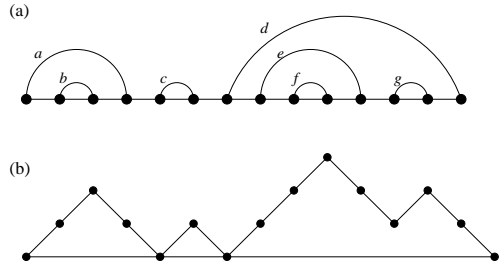}
\end{center}
\caption{\small{The graphical representation of a Catalan pair in
terms of a noncrossing matching, and the associated Dyck
path.}\label{esempio}}\vspace{-10pt}
\end{figure}

\begin{example}\label{xx}
{\em Let $X=\{ a,b,c,d,e,f,g\}$, and let $S$ and
$R$ be defined as follows:

\bigskip

$S=\{ (b,a),(f,e),(f,d),(e,d),(g,d)\}$

\smallskip

$R=\{ (a,c),(a,d),(a,e),(a,f),(a,g),(b,c),(b,d),(b,e),(b,f),(b,g),\\
\phantom{\indent R=\{}(c,d),(c,e),(c,f),(c,g),(e,g),(f,g)\} .$

\bigskip

It is easy to check that $(S,R)$ is a Catalan pair of size 7 on $X$, which can be represented as in Figure
\ref{esempio}~(a).}
\end{example}

An equivalent way to represent perfect noncrossing matchings is to
use Dyck paths: just interpret the leftmost element of an arch as
an up step and the rightmost one as a down step. For instance, the
matching represented in Figure \ref{esempio}~(a) corresponds to
the Dyck path depicted in Figure \ref{esempio}~(b). Coming back to
Catalan pairs, the relations $S$ and $R$ are suitably interpreted
using the notion of a tunnel. A \emph{tunnel} in a Dyck path
\cite{E} is a horizontal segment joining the midpoints of an up
step and a down step, remaining below the path and not
intersecting the path anywhere else. Now define $S$ and $R$ on the
set $X$ of the tunnels of a Dyck paths by declaring, for any
$x,y\in X$:
\begin{itemize}
\item $xSy$ when $x$ lies above $y$;

\item $xRy$ when $x$ is completely on the left of $y$.
\end{itemize}

See again Figure \ref{esempio} for an example illustrating the
above definition.

\subsection{Plane trees}

Let $\mathcal{T}_n$ be the set of plane trees having $n$ edges. It
is known that, the number of the elements of the set
$\mathcal{T}_n$ is the $n$th Catalan number. We say that a node
$b$ is a \emph{descendant} of a node $a$ when $b$ belongs to the
subtree of root $a$. In this situation, we also say that $a$ is an
\emph{ancestor} of $b$. For any two nodes $b$ and $c$, we define
their \emph{minimum common ancestor} to be the root of the minimum
subtree containing both $b$ and $c$. Finally, we will say that $b$
lies \emph{on the left of} $c$ when, called $a$ the minimum common
ancestor of $b$ and $c$, $a \notin \{b,c\}$ and $b$ is on the left
of $c$.

\begin{figure}[!htb]
\begin{center}
\centerline{\hbox{\psfig{figure=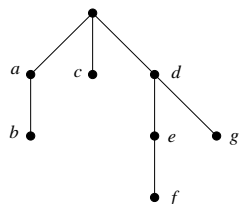,width=1.2in,clip=}}}
\caption{\small{The plane tree corresponding to the Catalan pair
represented in Figure~\ref{esempio}.}
\label{planetree}}\vspace{-30pt}
\end{center}
\end{figure}

Given $t\in \mathcal{T}_n$, let $X$ denote the set of nodes of $t$
other than the root. Define two relations $S$ and $R$ on $X$ as
follows:
\begin{itemize}
\item $xSy$ when $x$ is a descendant of $y$;

\item $xRy$ when $x$ lies on the left of $y$.
\end{itemize}
Then the pair $(S,R)$ is indeed a Catalan pair on $X$, and it
induces the well known (see \cite{St1}) bijection between plane trees and Dyck
paths. Figure~\ref{planetree} depicts the plane tree corresponding
to the Catalan pair $(S,R)$ considered in Example~\ref{xx}.

\subsection{Pattern avoiding permutations}
Let $n,m$ be two positive integers with $m\leq n$, and let $\pi
=\pi (1)\cdots \pi (n)\in S_n$ and $\nu =\nu (1)\cdots \nu (m)\in
S_m$. We say that $\pi$ {\em contains} the pattern $\nu$ if there
exist indices $i_1 <i_2 <\ldots <i_m$ such that $(\pi ({i_1 }),\pi
({i_2 }),\ldots ,\pi ({i_m }))$ is in the same relative order as
$(\nu (1),\ldots ,\nu(m))$. If $\pi$ does not contain $\nu$, we
say that $\pi$ is {\em $\nu$-avoiding}. See \cite{B} for plenty of
information on pattern avoiding permutations. For instance, if
$\nu =123$, then $\pi =524316$ contains $\nu$, while $\pi =632541$
is $\nu$-avoiding.

We denote by $S_n (\nu )$ the set of $\nu$-avoiding permutations
of $S_n$. It is known that, for each pattern $\nu \in S_3$, $|S_n
(\nu )|=C_n$ (see, for instance, \cite{B}).

\bigskip

It is possible to give a description of the classes of
312-avoiding permutations and 321-avoiding permutations in terms
of Catalan pairs. In this way, we are able to determine a description of the class
$S_n(\nu)$ in terms of Catalan pair, for any $\nu \in S_3$. From
the interpretation of $S_n(312)$ follows the interpretation of
$S_n(231)$ by inverse and the interpretation of $S_n(213)$ by
symmetry. From the interpretation of $S_n(321)$ follows the
interpretation of $S_n(123)$ by symmetry and from the
interpretation of $S_n(231)$ follows the interpretation of
$S_n(132)$ by symmetry.
\subsubsection{312-avoiding permutations}
Let $X=\{ 1,2,\ldots ,n\}$, for every permutation $\pi \in S_n$ we
define the following relations $S$ and $R$ on $X$:
\begin{itemize}
\item $iSj$ when $i<j$ and $(i,j)$ is an inversion in $\pi$ (i.e. $\pi(i) > \pi(j)$);

\item $iRj$ when $i<j$ and $(i,j)$ is a noninversion in $\pi$ (i.e $\pi(i) < \pi(j)$).
\end{itemize}

\begin{prop} The permutation $\pi \in S_n$ is 312-avoiding if and
only if $(S,R)$ is a Catalan pair of size $n$.
\end{prop}

\emph{Proof.}\quad The axioms (i) to (iii) in the definition of a
Catalan pair are satisfied by $(S,R)$ for any permutation $\pi$,
as the reader can easily check. Moreover, $\pi$ is 312-avoiding if
and only if, given any three positive integers $i<j<k$, it can
never happen that both $(i,j)$ and $(i,k)$ are inversions and
$(j,k)$ is a noninversion. This happens if and only if $S\circ R$
and $S$ are disjoint. But, from the above definitions of $S$ and
$R$, it must be $S\circ R\subseteq R\cup S$, hence $S\circ
R\subseteq R$. The axiom (iv) in the definition of a Catalan pair
is satisfied by $(S,R)$.\cvd
\\
\noindent \emph{Example.} We consider the following 312-avoiding permutation $\pi$:
\begin{displaymath}
\pi = {1 \ 2 \ 3 \ 4 \ 5 \ 6 \choose 2 \ 1 \ 3 \ 5 \ 6 \ 4}
\end{displaymath}
\noindent This configuration defines the following Catalan pair $(S,R)$:
\begin{description}
\item $S=\{(1,2),(4,6),(5,6)\}$ \ ;
\item $R=\{(1,3),(1,4),(1,5),(1,6),(2,3),(2,4),(2,5),(2,6),\\
(3,4),(3,5),(3,6),(4,5)\}$.\hspace{\stretch{1}} $\Box$
\end{description}

The present interpretation of 312-avoiding permutations
can be connected with the previous ones using Dyck paths and
perfect noncrossing matchings, giving rise to a very well-known
bijection, whose origin is very hard to be traced back (see, for
instance, \cite{P}). We leave all the details to the interested
reader.

\subsubsection{321-avoiding permutations}
\begin{figure}[!htb]
\begin{center}
\epsfig{file=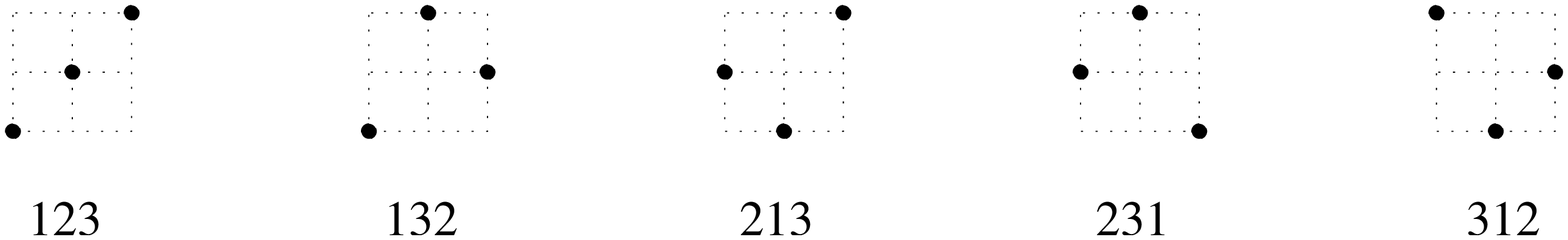,width=4in,clip=} \caption{\small{A
Graphical representation of $S_3(321)$ on the Cartesian plane.}
\label{plane}}\vspace{-10pt}
\end{center}
\end{figure}

Each permutation $\pi = (\pi(1)...\pi(n)) \in S_n$ can be
naturally represented on the Cartesian plane. In particular, each
element $\pi(i)$ of the permutation is the point $(i,\pi(i))$ on
the Cartesian plane,
for any $1 \leq i \leq n$.\\
\noindent For example, we represent the permutations of $S_3(321)$
on the Cartesian plane as in Figure \ref{plane}:\\

Let $X = \{a_1,a_2,..,a_n\}$ be the set of the points on the
Cartesian plane representing a permutation $\pi \in
S_n(321)$. If $x,y \in X$, where $x=(x_1,x_2)$ and $y=(y_1,y_2)$, we set $x \prec y$ when
$x_1 < y_1$ and $x \lhd y$ when $x_2 < y_2$.
%For any $a_i \in X$, $l(a_i)$ and $b(a_i)$ are the
%\emph{extensions} of the point $a_i$ as in Figure \ref{ex321}.\\
%\begin{figure}[!htb]
%\begin{center}
%\epsfig{file=perdue.eps,width=1in,clip=} \caption{A graphical
%representation of the extensions of the point $a_i \in X$.
%\label{ex321}}\vspace{-15pt}
%\end{center}
%\end{figure}

\begin{figure}[htb]
\begin{center}
\epsfig{file=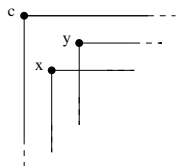,width=0.9in,clip=} \caption{\small{A
graphical representation of the cover of $(x,y)$.}
\label{cer}}\vspace{-15pt}
\end{center}
\end{figure}

Using this representation, we can define the following relations
on the set $X$. Given $x,y \in X$, a \emph{cover} of
$\{x,y\}$ is any point $c$ of $X$
having the following properties (see
Figure \ref{cer}):
\begin{itemize}
\item $x \lhd c$ and $y \lhd c$ ; \item $c \prec x$ and $c \prec
y$.
\end{itemize}
%\begin{itemize}
%\item{-} $\pi(c) > \pi(x) \ \ and \ \ \pi(c) > \pi(y)$ \ , \item{-}
%$l(c) \cap b(x) = \emptyset \ \ and \ \ l(x) \cap b(c) =
%\emptyset$ \ , \item{-} $l(c) \cap b(y) = \emptyset \ \ and \ \
%l(y) \cap b(c) = \emptyset$ \ .
%\end{itemize}
%Roughly speaking, a cover of $(x,y)$ is any element of $X$
%such that its extensions cover both the extensions of the points
%$x$ and $y$, without intersecting them.

\noindent For any $x,y \in X$, we say that:

\begin{itemize}
\item[]$\bullet$ $xRy$ when there is no cover of
$\{x,y\}$, $x \lhd y$ and $x \prec y$ \ ;
\item[]$\bullet$ $xSy$ when $(x,y) \notin \overline{R}$ and $x \prec y$.

%\begin{displaymath} \hspace{-125pt}
%\bullet \ xSy \ \textrm{when} \left\{ \begin{array}{ll}
%\textrm{$\pi(x) < \pi(y)$ \ \ with }  \\
%\textrm{$l(x) \cap b(y) = \emptyset$  \ and \ $b(x) \cap l(y) = \emptyset$ \ \ \ \ \ \ \ \ \ (1)}\\
%\\
%\textrm{there is a cover of $(x,y)$ and}\\
%\textrm{$\pi(x) < \pi(y)$ with $l(x) \cap b(y) \neq \emptyset$.\ \ \ \ \ \ \ \ \ \ \ \ \ \ (2)}\\
%\end{array} \right.
%\end{displaymath}
\end{itemize}

\begin{figure}[!htb]
\begin{center}
\epsfig{file=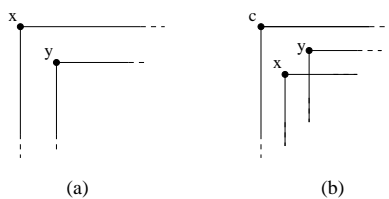,width=2in,clip=} \caption{\small{A
graphical representation of distinct cases of the relation $S$.}
\label{prt}}\vspace{-15pt}
\end{center}
\end{figure}

We can observe that the definition of the relation $S$ consists in
two distinct cases which are depicted in Figure \ref{prt}.

\begin{prop} $(S,R)$ is Catalan pair on the set $X$.
\end{prop}

\emph{Proof.}\quad The axioms (i) to (iii) in the definition of a
Catalan pair are satisfied by $(S,R)$, as the reader can easily check.

\begin{figure}[htb]
\begin{center}
\epsfig{file=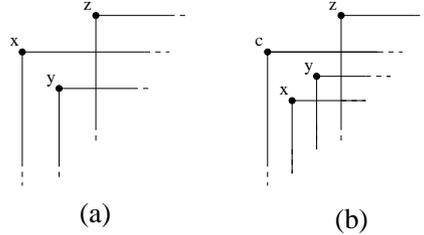,width=2.2in,clip=} \caption{\small{A
graphical representations of the configuration $xSyRz$.}
\label{prof}}\vspace{-15pt}
\end{center}
\end{figure}

Let $x,y,z \in X$ such that $xSy$ and $yRz$. From the above definition of the
relation $S$ it follows that the configuration $xSyRz$ can be represented by
two distinct cases, as in Figure \ref{prof}.\\

Both configurations (a) and (b) satisfy the relation $xRz$, hence
$S\circ R\subseteq R$. The axiom (iv) in the definition of a
Catalan pair is satisfied by $(S,R)$. \cvd

\begin{figure}[htb]
\begin{center}
\epsfig{file=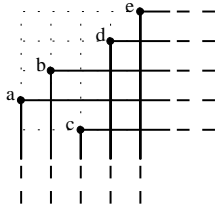,width=1.1in,clip=} \caption{\small{A
graphical representation of $\pi$.} \label{sds}}\vspace{-15pt}
\end{center}
\end{figure}

\emph{Example.} We consider the following 321-avoiding permutation $\pi$:
\begin{displaymath}
\pi = {1 \ 2 \ 3 \ 4 \ 5 \choose 2 \ 3 \ 1 \ 4 \ 5}
\end{displaymath}
\\
\noindent The Figure \ref{sds} shows the graphical representation of $\pi$.\\
\\
\noindent This configuration defines the following Catalan pair $(S,R)$ on the $X=\{a,b,c,d,e\}$:
\begin{description}
\item $S=\{(a,c),(b,c)\}$ \ ;
\item $R=\{(a,b),(a,d),(a,e),(b,d),(b,e),(c,d),(c,e),(d,e)\}$. \hspace{\stretch{1}} $\Box$\\
\end{description}

\subsection{Sequences of integers counted by Catalan numbers}
There are many classes of integers sequences satisfying special
constraints which are enumerated by Catalan numbers. In
\cite{St1} we can find some examples, among which we focus on the
following ones: \begin{itemize} \item[(1)] the set of sequences $a_1 \
a_2 \ ..\ a_n$ of integers with $i \leq a_i \leq n$ and such that
if $i \leq j \leq a_i$, then $a_j \leq a_i$. \item[(2)] the set of
sequences $1 \leq a_1 \leq a_2 \leq .. \leq a_n \leq n$ of
integers with exactly one fixed point, i.e., exactly one index $1
\leq f \leq n$ for which $a_f=f$.
\end{itemize}
Let us study these two classes separately:
\begin{itemize}
\item[(1)] Let $X = \{a_1,a_2,..,a_n\}$ be the set of integers which form the sequence. For instance, for $n=3$ we have the following five
sequences on $X=\{1,2,3\}$:
\begin{figure}[htb]
\begin{center}
\epsfig{file=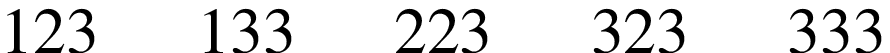,width=2.1in,clip=} \vspace{-10pt}
\end{center}
\end{figure}

We can define the relations $S$ and $R$ on the set $X$, as follows.\\
\noindent For any $a_i, a_j \in X$, with $1 \leq i,j \leq n$:
\begin{itemize}
\item[$\bullet$] $a_i R a_j$ when $i < j$ and $a_i < a_j$ ;
\item[$\bullet$] $a_i S a_j$ when $j < i$ and $a_i \leq a_j$.
\end{itemize}

\begin{prop} $(S,R)$ is Catalan pair on the set $X$.
\end{prop}

\emph{Proof.}\quad The axioms (i) to (iii) in the definition of a
Catalan pair are satisfied by $(S,R)$, as the reader can easily check.

Let $a_i, a_j, a_k \in X$ with $i < j < k$ such that $a_jSa_i$ and
$a_iRa_k$. From the above definitions of the relations $S$ and $R$
it follows that the relation $a_jSa_i$ is satisfied when $i < j$
and $a_j \leq a_i$, while the relation $a_iRa_k$ is satisfied when
$i < k$ and $a_i < a_k$. Then $a_j < a_k$ and since $j < k$ it
follows that the relation $a_jRa_k$ is satisfied, hence $S\circ
R\subseteq R$. The axiom (iv) in the definition of a Catalan pair
is satisfied by $(S,R)$.\cvd

\emph{Example.} We consider the following sequence:
\begin{displaymath}
a_1 \ a_2 \ a_3 \ a_4 \ a_5 \ a_6 \choose 5 \  \ 2 \  \ 4 \  \ 4 \  \ 5 \  \ 6
\end{displaymath}
\noindent This configuration defines the following Catalan pair $(S,R)$ on the set $X=\{a_1,a_2,..,a_6 \}$:
\begin{description}
\item $S=\{(a_2,a_1),(a_3,a_1),(a_4,a_1),(a_5,a_1),(a_4,a_3)\}$ \ ;
\item $R=\{(a_1,a_6),(a_2,a_3),(a_2,a_4),(a_2,a_5),(a_2,a_6),(a_3,a_5),(a_3,a_6),(a_4,a_5),\\
(a_4,a_6),(a_5,a_6)\}$.\hspace{\stretch{1}} $\Box$
\end{description}

\item[(2)]As above, let $X = \{a_1,a_2,..,a_n\}$ be the set of
integers which form the sequence. For instance, for $n=3$ we have
the following five sequences on $X=\{1,2,3\}$:
\begin{figure}[htb]
\begin{center}
\epsfig{file=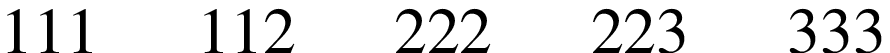,width=2.1in,clip=}\vspace{-10pt}
\end{center}
\end{figure}

Let $f$ be the fixed point of $X$, i.e. the index such that $a_f=f$; for any index $y \leq f$ we consider
the integer $a_y'=a_y-y$ and for any index $z$ with $f < z \leq n$ we consider the integer $a_z'=z-a_z$.
In this way, for any sequence $a_1 \ a_2 \ .. \ a_n$ we have the corresponding sequence $a_1' \ a_2' \ .. \ a_n'$.\\
Using this representation, we can define the following relations on the set $X$. Let $i,j$ be the indexes of the
integers $a_i, a_j \in X$, we can describe the following cases:
\begin{itemize}
\item [$\bullet$] If $i,j \leq f$ then:
\begin{description}
\item[] \ \ \ \ \ \ \ $a_iSa_j$ when \ $i<j$ \ , \ $a_i'>a_j'$ \
and \ $j=min\{k:i<k \leq f, \ a_k'=a_j'\}$ ; \item
\begin{displaymath} \hspace{-65pt} \textrm{$a_iRa_j$} \
\textrm{when} \left\{
\begin{array}{ll}
\textrm{$i<j$ \ , \ $a_i'>a_j'$ \ and }  \\
\textrm{$\exists w$ with $i<w<j$ such that $a_w'=a_j'$ ; \ \ \ \ (1)}\\
\\
\\
\textrm{$i<j$ \ and \ $a_i' \leq a_j'$.\ \ \ \ \ \ \ \ \ \ \ \ \ \ \ \ \ \ \ (2)}\\

\end{array} \right.
\end{displaymath}
\end{description}
Roughly speaking, we say that $a_iSa_j$ if $i<j$ and $a_i'>a_j'$
where $a_j'$ is the first integer in the sequence
$a_{i+1}'a_{i+2}'\dots a_f'$ having that value and we say that
$a_iRa_j$ for (1) if $i<j$ and $a_i'>a_j'$ where $a_j'$ is not the
first integer in the sequence $a_{i+1}'a_{i+2}'\dots a_f'$ having
that value, but exists an integer $a_w'$ with $i<w<j$ such that
$a_w'=a_j'$. \item[$\bullet$] If $i \leq f < j$ then $a_iRa_j$.
\item[$\bullet$] If $i,j>f$ then:
\begin{description}
\item \ \ \ \ \ \ \ $a_iS^{-1}a_j$ ($a_jSa_i$) when \ $i<j$ \ , \
$a_i'<a_j'$ \ and \ $i=max\{k:f<k<j \ , \ a_k'=a_i'\}$~; \item
\begin{displaymath} \hspace{-75pt} \textrm{$a_iRa_j$} \
\textrm{when} \left\{
\begin{array}{ll}
\textrm{$i<j$ \ , \ $a_i'<a_j'$ \ and }  \\
\textrm{$\exists w$ with $i<w<j$ such that $a_w'=a_i'$ ; \ \ \ \ (1)}\\
\\
\\
\textrm{$i<j$ \ and \ $a_i' \geq a_j'$.\ \ \ \ \ \ \ \ \ \ \ \ \ \ \ \ \ \ \ (2)}\\

\end{array} \right.
\end{displaymath}
\end{description}
Roughly speaking, we say that $a_iS^{-1}a_j$ (or equivalently
$a_jSa_i$) if $i<j$ and $a_i'<a_j'$ where $a_i'$ is the last
integer in the sequence $a_{f+1}'a_{f+2}'\dots a_{j-1}'$ having
that value and we say that $a_iRa_j$ for (1) if $i<j$ and
$a_i'<a_j'$ where $a_i'$ is not the last integer in the sequence
$a_{f+1}'a_{f+2}'\dots a_{j-1}'$ having that value, but exists an
integer $a_w'$ with $i<w<j$ such that $a_w'=a_i'$.
\end{itemize}
\begin{prop} \label{sequen}
Let $a_1' \ a_2' \ .. \ a_n'$ be the corresponding sequence of $a_1 \ a_2 \ .. \ a_n$ of integers with exactly
one fixed point $f=a_f$. If there are indexes $x<y<z<f$ with $a_x', a_z'>a_y'$ and $a_x'>a_z'$ then there is
an index $w$ with $x<w<y$ such that $a_w'=a_z'$.\cvd
\end{prop}
\begin{prop} $(S,R)$ is Catalan pair on the set $X$.
\end{prop}

\emph{Proof.}\quad The axioms (i) to (iii) in the definition of a
Catalan pair are satisfied by $(S,R)$, as the reader can easily check.

Let $a_1 \ a_2 \ .. \ a_n$ be the sequence of integers with exactly one fixed point $f=a_f$ and
let $a_1' \ a_2' \ .. \ a_n'$ be the corresponding sequence of $a_1 \ a_2 \ .. \ a_n$.\\
\noindent Let $a_i, a_j, a_z \in X$ such that $a_iSa_j$ and $a_jRa_z$.

Consider the case $i<j<z<f$. From the above definitions of the
relations $S$ and $R$ it follows that the relation $a_iSa_j$ is
satisfied when $i < j$ ,$a_i' > a_j'$ and $j=min\{k:i<k \leq f, \
a_k'=a_j'\}$, while the relation $a_jRa_z$ can be satisfied by two
distinct cases: the case (1) or the case (2). Suppose that
$a_jRa_z$ is satisfied by the case (1), when $j < z$ , $a_j' >
a_z'$ and there is an index $w$ with $j<w<z$ such that
$a_w'=a_z'$. Then $a_i' > a_z'$ and since $i<j$ it follows that
there is an index $w$ with $i<w<z$ such that $a_w'=a_z'$, hence
the relation $a_iRa_z$ is satisfied by the case (1) of $R$.

Now, suppose that $a_jRa_z$ is satisfied by the case (2) of the
relation $R$, when $j<z$ and $a_j' \leq a_z'$. If $a_j' = a_z'$
then $a_i' > a_z'$ and since $i<j<z$ it follows that the relation
$a_iRa_z$ is satisfied by the case (1) of $R$. If $a_j'<a_z'$ , it
must be either $a_i' \leq a_z'$ or $a_i'>a_z'$. If $a_i' \leq
a_z'$~, since $i<z$ it follows that the relation $a_iRa_z$ is
satisfied by the case (2) of $R$. If $a_i'>a_z'$~, from the
Proposition \ref{sequen} it follows that there is an index $w$
with $i<w<j$ such that $a_w'=a_z'$ , since $i<j<z$ , the relation
$a_iRa_z$ is satisfied by the case (1) of $R$. Thus, we can
conclude that, in every case, $a_iRa_z$, hence $S\circ R\subseteq
R$ for $i<j<z<f$.

The case $i,j,z>f$ can be treated analogously and the case
$i,j<f<z$ is obvious, hence the axiom (iv) in the definition of a
Catalan pair is satisfied by $(S,R)$.\cvd

\emph{Example.} We consider the following sequence:
\begin{displaymath}
a_1 \ \ a_2 \ \ a_3 \ \ a_4 \ \ a_5 \ \ a_6 \ \ a_7 \ \ a_8 \choose 2 \ \ \ 4 \ \ \ 4 \ \ \ 5 \ \ \ 5 \ \ \ 5 \ \ \ 6 \ \ \ 6
\end{displaymath}
\noindent The corresponding sequence of $a_1 \ a_2 \ .. \ a_n$ is
the sequence $a_1' \ a_2' \ .. \ a_n'$ defined as follows:
\begin{displaymath}
a_1' \ \ a_2' \ \ a_3' \ \ a_4' \ \ a_5' \ \ a_6' \ \ a_7' \ \ a_8' \choose 1 \ \ \ 2 \ \ \ 1 \ \ \ 1 \ \ \ 0 \ \ \ 1 \ \ \ 1 \ \ \ 2
\end{displaymath}
\noindent This configuration defines the following Catalan pair $(S,R)$ on the set $X=\{a_1,a_2,..,a_8 \}$:
\begin{description}
\item $S=\{(a_1,a_5),(a_2,a_3),(a_2,a_5),(a_3,a_5),(a_4,a_5),(a_8,a_7)\}$ \ ;
\item $R=\{(a_1,a_2),(a_1,a_3),(a_1,a_4),(a_1,a_6),(a_1,a_7),(a_1,a_8),(a_2,a_4),(a_2,a_6),\\
(a_2,a_7),(a_2,a_8),(a_3,a_4),(a_3,a_6),(a_3,a_7),(a_3,a_8),(a_4,a_6),(a_4,a_7),\\
(a_4,a_8),(a_5,a_6),(a_5,a_7),(a_5,a_8),(a_6,a_7),(a_6,a_8)\}$.\hspace{\stretch{1}} $\Box$
\end{description}
\end{itemize}
\subsection{Staircase shape}
\begin{figure}[htb]
\begin{center}
\epsfig{file=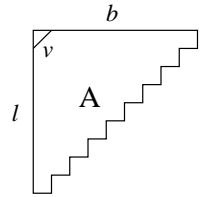,width=1in,clip=} \caption{\small{A
graphical representation of staircase shape A.}
\label{scala}}\vspace{-15pt}
\end{center}
\end{figure}

A \emph{staircase shape} (see \cite{St2}) $A$ is depicted in
Figure \ref{scala}. In particular $|b|=|l|=n$ which is the
\emph{size} of the staircase shape having $n$ steps of the form
{\Huge $\lrcorner$}. Two staircase shapes of size $n$ are said to
be different one the other when they are
divided into exactly $n$ rectangles in two different ways. The following Figure~\ref{caso} shows the case $n=3$.\\
%an object formed by two
%perpendicular segments $b(A)$ and $l(A)$ such that $|b(A)|=|l(A)|$
%and by a sequence of segments with unitary length which are
%perpendicular two by two (see Figure \ref{scala}). In particular,
%the sequence of segments with unitary length made up one
%horizontal segment followed by one vertical segment is called
%\emph{step}. Let $v(A)$ be the angle between the segments $b(A)$
%and $l(A)$.

%We say that, the staircase shape $A$ has size $n$ if
%$|b(A)|=|l(A)|=n$ and there are $n$ rectangles which lite $A$. It
%is known that, the number of staircase shapes of size $n$ is the
%$n$st Catalan number (see \cite{St2}). The following Figure~\ref{caso} shows the case $n=3$.\\
\begin{figure}[!htb]
\begin{center}
\epsfig{file=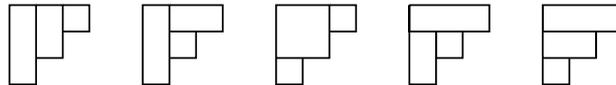,width=3.2in,clip=} \caption{\small{A
graphical representation of all staircase shapes of size 3.}
\label{caso}}\vspace{-15pt}
\end{center}
\end{figure}
\begin{prop}\label{sca}

Each staircase shape of size $n$, with $n$ rectangles, has $n$
steps and each step belong to one and only one rectangle.\cvd
\end{prop}

\begin{figure}[htb]
\begin{center}
\epsfig{file=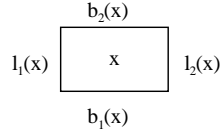,width=1.1in,clip=} \caption{\small{A
graphical representation of the labelled sides of $x$.}
\label{etic}}\vspace{-15pt}
\end{center}
\end{figure}

Let $X(A)$ be the set of the rectangles which tiles the staircase
shape $A$. For any $x \in X(A)$, the sides of the rectangle $x$
are labelled as in Figure \ref{etic}.\\

\begin{figure}[htb]
\begin{center}
\epsfig{file=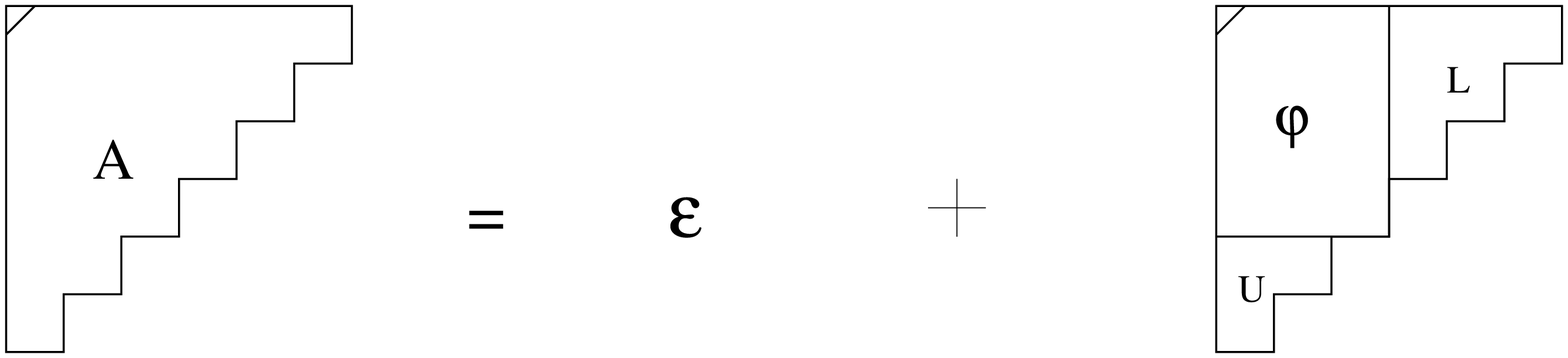,width=3in,clip=}\caption{\small{A
graphical representation of the unique decomposition of staircase
shape $A$.} \label{decompos}}\vspace{-15pt}
\end{center}
\end{figure}

\begin{prop}\label{deco}
Each staircase shape $A$ of size $n$ admits the unique
decomposition in Figure~\ref{decompos}, where:
\begin{description}
\item[$\bullet$] $\varepsilon$ is the empty staircase shape;
\item[$\bullet$] $\varphi \in X(A)$ is the junction rectangle
containing the angle $v$ ; \item[$\bullet$] $L$ and $U$ are
staircase shapes of size $m$ and $m'$ respectively , with $m,m' <
n$.
\end{description}
\end{prop}

\emph{Proof.}\quad Let $A$ be a staircase shape of size $n$, with
$n \geq 1$. Just locate the junction rectangle $\varphi \in X(A)$
to determine the decomposition of $A$. The junction rectangle of
the configuration $A$ is the rectangle of the set $X(A)$ which
contains the angle $v$.

From Proposition \ref{sca}, it follows that $|l_1(\varphi)|=k+1$
and $|b_1(\varphi)|=n-k$ with $0 \leq k \leq n-1$. By the way, the
staircase shape $L$ of size $k$, lies completely on the right of
$l_1(\varphi)$, while the staircase shape $U$ of size $n-(k+1)$,
lies under $b_1(\varphi)$.

The uniqueness of the decomposition of $A$ is related to the
existence of only one junction rectangle $\varphi \in X(A)$.\cvd

At this point, we can define the following relations on the set
$X$ of rectangles which tile a staircase shape. For any $x,y \in
X$, we set:
\begin{itemize}
\item $xSy$ when $l_2(x)$ (or the extension of $l_2(x)$) intersects $b_1(y)$ ;
\item $xRy$ when $x$ is completely on the left of $l_1(y)$ (or the extension of $l_1(y)$).
\end{itemize}

\begin{prop} $(S,R)$ is Catalan pair on the set $X$.
\end{prop}

\emph{Proof.}\quad The axiom (i) in the definition of a
Catalan pair is satisfied by $(S,R)$, as the reader can easily check.

Let $x,y \in X$ be two distinct rectangles of a staircase shape
$A$. From Proposition \ref{deco}, it follows that the staircase
shape $A$ admits a unique decomposition, then we can consider the
following cases:
\begin{description}
\item a) $x=\varphi$,\\
it must be either $y \in L$ or $y \in U$. If $y \in L$ then $xRy$,
otherwise if $y \in U$ then $ySx$;
\item b) $y=\varphi$,\\
it must either $x \in L$ or $x \in U$. If $x \in L$ then $yRx$,
otherwise if $x \in U$ then $xSy$.
\end{description}
Only one relation between $x$ and $y$ can hold in $A$, hence the
axioms (ii) and (iii) in the definition of a Catalan pair are
satisfied by $(S,R)$.

Let $x,y,z \in X$ be two distinct rectangles of a staircase shape
$A$, such that $xSy$ and $yRz$. From Proposition \ref{deco}, it
follows that the staircase shape $A$ admits a unique
decomposition. The most interesting case is when $y=\varphi$.
Since $xSy$ and $yRz$, it follows that $x \in U$ and $z \in L$,
hence $xRz$. The axiom (iv) in the definition of a Catalan pair is
satisfied by $(S,R)$. \cvd

\begin{figure}[!htb]
\begin{center}
\epsfig{file=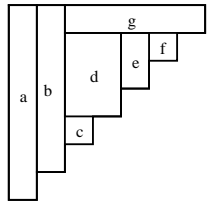,width=1.041in,clip=} \caption{\small{An
example of staircase shape of size 7.} \label{rog}}\vspace{-15pt}
\end{center}
\end{figure}

\emph{Example}. \quad Let $X=\{a,b,c,d,e,f,g\}$ be the set of the
rectangles which tile the
staircase shape of size 7 represented in Figure \ref{rog}.\\

\noindent This configuration defines the following Catalan pair $(S,R)$ on the set $X=\{a,b,c,d,e,f,g\}$:
\begin{description}
\item $S=\{(c,d),(c,g),(d,g),(e,g),(f,g)\}$ ;
\item $R=\{(a,b),(a,c),(a,d),(a,e),(a,f),(a,g),(b,c),(b,d),(b,g),(b,e),(b,f),\\
(c,e),(c,f),(d,e),(d,f),(e,f)\}$. \hspace{\stretch{1}} $\Box$
\end{description}

\section{Further work}

At the end of the paper, we would like to take into considerations the Catalan structures which do not admit a recursive decomposition as (\ref{deccs}). Among these ones, the most popular are perhaps the parallelogram polyominoes, the $2$-colored Motzkin paths,
the binary trees \cite{St1}.

By means of the following example, concerning the class of parallelogram polyominoes, we show that it is possible to adapt our
method to include also these  "more complex'' combinatorial structures.

\begin{figure}[!htb]
\begin{center}
\epsfig{file=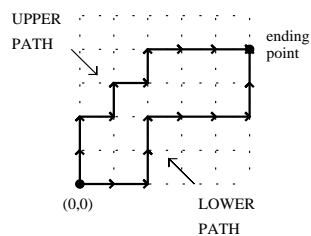,width=1.6in,clip=} \caption{\small{A
parallelogram polyomino with semi-perimeter 9.}
\label{poli}}\vspace{-15pt}
\end{center}
\end{figure}

A parallelogram polyomino with semi-perimeter $n+1$ is defined by
two distinct non intersecting lattice paths of length $n+1$
beginning in $(0,0)$ and using only horizontal and vertical unit
steps. These paths, called the upper and the lower path,
respectively, meet in only two points, the beginning point and the
ending point of both, see Figure \ref{poli}.\\

\begin{figure}[htb]
\begin{center}
\epsfig{file=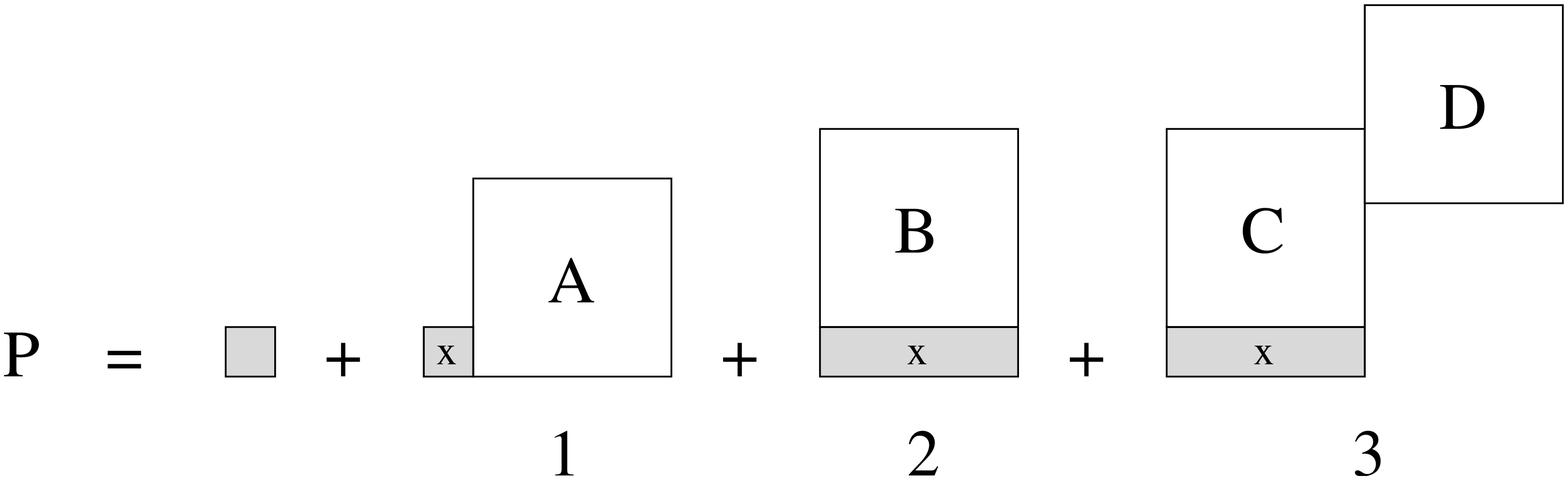,width=4in,clip=}
\caption{\small{Recursive decomposition of the parallelogram
polyominoes.} \label{decpoliom}}\vspace{-15pt}
\end{center}
\end{figure}

The class $\mathcal{P}$ of the parallelogram polyominoes is counted by the Catalan numbers according to the semi-perimeter, but $\mathcal{P}$ does not admit a decomposition as (\ref{deccs}). For this class, a simpler decomposition is given by
that depicted in Figure \ref{decpoliom}, where $x$ is a single element of unitary size belonging to the base set
$X$ and
$A,B,C,D$ are parallelogram polyominoes of lower size.\\

%\begin{figure}[htb]
%\begin{center}
%\epsfig{file=decpoliexbis.eps,width=3.5in,clip=}
%\caption{\small{Recursive decomposition of the parallelogram
%polyomino in Figure \ref{poli}.}
%\label{decpoliomex}}\vspace{-15pt}
%\end{center}
%\end{figure}

Figure~\ref{decpoliom} shows that we have three distinct
operations, denoted 1,2 and 3 respectively, which can be applied
on the class of parallelogram polyominoes. As we did for
decomposition (\ref{decc}), also in this case we can recursively
define Catalan pairs $(S,R)$ on the class of parallelogram
polyominoes.

In particular, if $P$ is a parallelogram polyomino, and it is not
the single cell, then it can be uniquely decomposed according to
1, 2 or 3:

\begin{enumerate}
\item if the last operation applied on $P$ is operation $1$, let $(S_A,R_A)$ be the Catalan pair on the base set for $A$, then $S$ and $R$ are defined as follows:
$$ \, S=S_A, \quad R=R_A \cup \{(x,a): a \in A \} \, . $$
\item if the last operation applied on $P$ is operation $2$, let $(S_B,R_B)$ be the Catalan pair on the base set for $B$, then $S$ and $R$ are defined as follows:
$$ \, S=S_B \cup \{(b,x): b \in B\}, \quad R=R_B. \, $$
\item if the last operation applied on $P$ is operation $3$, let $(S_C,R_C)$, $(S_D,R_D)$ be the Catalan pairs on $C$, $D$, respectively, then $S$ and $R$ are defined as follows:
$$ \, S=S_C \cup S_D \cup \{(c,x): c \in C \}, \quad R=R_C \cup R_D \cup \{(c,d): c \in C, d \in D \} \cup \{(x,d): d \in D \} \, $$
\end{enumerate}

The construction of the base set $X$ follows the method previously described and is straightforward.

Applying the previous technique it is then possible to automatically determine the Catalan pairs
associated with all the structures which satisfy a decomposition like that in Figure~\ref{decpoliom}.

\end{document}